\documentstyle[12pt]{article}

\newcommand{\bm}[1]{\mbox{\bf #1}}
\makeatletter
\def\lsim{\mathrel{\mathpalette\subsim@align<}}
\def\gsim{\mathrel{\mathpalette\subsim@align>}}
\def\subsim@align#1#2{\lower.6ex\vbox{\baselineskip\z@skip\lineskip\z@
\ialign{$\m@th#1\hfil##\hfil$\crcr#2\crcr\sim\crcr}}}
\makeatother
\begin{document}

\pagestyle{empty}
\begin{titlepage}

\title{Realistic shell model calculation
of $2\nu\beta\beta$ nuclear matrix elements\\
and role of shell structure in intermediate states}

\author{H. Nakada\\
{\small\it Department of Physics, Faculty of Science,
Chiba University,}\\
{\small\it Yayoi-cho 1-33, Inage-ku, Chiba 263, Japan}\\
T. Sebe\\
{\small\it Department of Applied Physics,
College of Engineering, Hosei University,}\\
{\small\it Kajino-cho 3-7-2, Koganei, Tokyo 184, Japan}\\
and\\
K. Muto\\
{\small\it Department of Physics, Faculty of Science,
Tokyo Institute of Technology,}\\
{\small\it Oh-okayama 2-12-1, Meguro-ku, Tokyo 152, Japan}}

\date{}

\maketitle
\thispagestyle{empty}

\begin{flushleft}
\end{flushleft}

\begin{abstract}
We discuss two conditions needed for correct computation
of $2\nu\beta\beta$ nuclear matrix-elements
within the realistic shell-model framework.
An algorithm in which intermediate states are treated
based on Whitehead's moment method is inspected,
by taking examples of the double GT$^+$ transitions
$\mbox{$^{36}$Ar}\rightarrow\mbox{$^{36}$S}$,
$\mbox{$^{54}$Fe}\rightarrow\mbox{$^{54}$Cr}$
and $\mbox{$^{58}$Ni}\rightarrow\mbox{$^{58}$Fe}$.
This algorithm yields rapid convergence
on the $2\nu\beta\beta$ matrix-elements,
even when neither relevant GT$^+$
nor GT$^-$ strength distribution is convergent.
A significant role of the shell structure is pointed out,
which makes the $2\nu\beta\beta$ matrix-elements
highly dominated by the low-lying intermediate states.
Experimental information of the low-lying GT$^\pm$ strengths
is strongly desired.
Half-lives of $T^{2\nu}_{1/2}({\rm EC}/{\rm EC};
\mbox{$^{36}$Ar}\rightarrow\mbox{$^{36}$S})
=1.7\times 10^{29}\mbox{yr}$,
$T^{2\nu}_{1/2}({\rm EC}/{\rm EC};
\mbox{$^{54}$Fe}\rightarrow\mbox{$^{54}$Cr})
=1.5\times 10^{27}\mbox{yr}$,
$T^{2\nu}_{1/2}({\rm EC}/{\rm EC};
\mbox{$^{58}$Ni}\rightarrow\mbox{$^{58}$Fe})
=6.1\times 10^{24}\mbox{yr}$
and $T^{2\nu}_{1/2}(\beta^+/{\rm EC};
\mbox{$^{58}$Ni}\rightarrow\mbox{$^{58}$Fe})
=8.6\times 10^{25}\mbox{yr}$
are obtained from the present realistic shell-model calculation
of the nuclear matrix-elements.
\end{abstract}

\end{titlepage}
\pagestyle{plain}
\setcounter{page}{1}

\section{Introduction}
\label{sec:intro}

The nuclear double-$\beta$ ($\beta\beta$) decay is a good probe
to investigate neutrino masses and some basic properties
of the weak interaction\cite{ref:DKT85,ref:Ej91,ref:Hax84}.
The $0\nu$ mode, if it exists, immediately indicates
a defect of the standard model.
On the other hand, the $2\nu$ mode occurs within the standard model.
This mode has a practical importance,
because it governs the lifetime
of the nuclei which concern $\beta\beta$ decay.
Moreover, the observation of the $2\nu\beta\beta$ decays\cite{ref:Ej91}
has revealed serious discrepancies
between earlier predictions and the observed half-lives.
It probably originated in a problem of the nuclear structure theories.
This problem may be crucial also to the $0\nu\beta\beta$ decays,
because some of the nuclear matrix-elements
associated with the $0\nu$ mode
have similarity to the $2\nu$ mode.

The $2\nu\beta\beta$ decay takes place
by two sequential Gamow-Teller (GT) transitions,
via virtual intermediate states.
It is not always possible
for the current nuclear structure theories
to give reliable predictions even on single-GT transitions,
since the GT transitions are sensitive to some details
of nuclear many-body wavefunctions.
The most promising approach to the GT transitions
seems to be realistic shell-model calculations.
It is yet difficult to apply realistic shell-model approaches
to the relatively heavy nuclei
where the $2\nu\beta\beta$ decays have been observed.
It may become possible in the near future, however,
to carry out a reliable shell-model calculation
in a few of these nuclei,
with assistance of the growing computer power.

Several realistic shell-model calculations have been reported
for the $\mbox{$^{48}$Ca}\rightarrow\mbox{$^{48}$Ti}$
$\beta^-\beta^-$ decay\cite{ref:ZBR90,ref:CPZ90,ref:HO89},
although it has not been observed.
In Ref.\cite{ref:ZBR90},
it has been pointed out that,
among a large number of possible intermediate states,
the lowest $1^+$ state gives a dominant contribution
to the $2\nu\beta\beta$ process.
In Ref.\cite{ref:CPZ90},
a method to handle the intermediate states has been developed
based on Whitehead's moment method\cite{ref:Wh80}.
Rapid convergence by this method has been shown
for the $^{48}$Ca decay.
The $^{48}$Ca case is, however, somewhat special,
as will be discussed later.
It is worth examining generality of these suggestions,
by taking other sample nuclei.

While most experimental efforts have been cast
on the double-$\beta^-$ ($\beta^-\beta^-$) decays,
the $\beta^+$ side is also important.
Since the electron capture (EC) and positron emission ($\beta^+$)
are possible in a single-GT$^+$ transition,
we have three cases in the double-GT$^+$ transitions;
EC/EC, $\beta^+$/EC and $\beta^+\beta^+$\cite{ref:DK92,ref:DK93}.
It has been pointed out that, in the $0\nu$ mode,
they seem to yield quite a different constraint
on the neutrino masses
and right-handed weak-current parameters\cite{ref:HMOK},
although the double-GT$^+$ transitions
are more difficult to be observed than the $\beta^-\beta^-$ cases.
From the viewpoint of the nuclear structure theories,
there is no essential distinction in the algorithm
to compute the $\beta\beta$ matrix-elements,
between the $\beta^-$ and the $\beta^+$ cases.
In the $A\lsim 60$ region,
where realistic shell-model wavefunctions are available,
the $\beta^+$ side delivers a few candidates for the $\beta\beta$ decay
in addition to $^{48}$Ca.
These candidates give us a good opportunity
to test the theoretical methods.
One important issue is practical treatment of the intermediate states.
Primarily focusing on this point,
we shall discuss in this article how to compute
the $2\nu\beta\beta$ nuclear matrix-elements
within the realistic shell-model framework,
by taking a few sample cases from the $\beta^+$ side;
the $\mbox{$^{36}$Ar}\rightarrow\mbox{$^{36}$S}$,
$\mbox{$^{54}$Fe}\rightarrow\mbox{$^{54}$Cr}$
and $\mbox{$^{58}$Ni}\rightarrow\mbox{$^{58}$Fe}$ decays,
where realistic shell-model wavefunctions are available
for the initial and final states\cite{ref:BW88,ref:NSO94}.
We will also discuss the key features to describing
the $2\nu\beta\beta$ nuclear matrix-elements.

\section{Shell model calculation
of $2\nu\beta\beta$ matrix elements}
\label{sec:comp}

In this section, we present a realistic shell-model calculation
of the $2\nu\beta\beta$ nuclear matrix-elements.
We discuss the conditions
required for treating the intermediate states.
The algorithm proposed in Ref.\cite{ref:CPZ90}
is argued in the light of these conditions,
and is re-examined for the examples
of the $\mbox{$^{36}$Ar}\rightarrow\mbox{$^{36}$S}$,
$\mbox{$^{54}$Fe}\rightarrow\mbox{$^{54}$Cr}$
and $\mbox{$^{58}$Ni}\rightarrow\mbox{$^{58}$Fe}$ decays.

With respect to the $^{36}$Ar decay,
a realistic shell-model hamiltonian has been established
in the {\em sd}-shell region;
the so-called USD interaction\cite{ref:BW88}.
The $^{54}$Fe and $^{58}$Ni nuclei
belong to the middle {\em pf}-shell region.
Though diagonalization of a hamiltonian in the full {\em pf}-shell
is not yet possible for the middle {\em pf}-shell nuclei,
a large-scale calculation in a loosely truncated space
is successful to describe spectroscopic properties\cite{ref:NSO94}.
The $k\leq 2$ configuration space, where $k$ represents
number of nucleons excited from $0f_{7/2}$
to ($0f_{5/2}$$1p_{3/2}$$1p_{1/2}$),
is adopted in this calculation,
together with the Kuo-Brown hamiltonian\cite{ref:KBpf}.
The $2\nu\beta\beta$ nuclear matrix-elements contain
a product of a GT$^+$ matrix-element from the initial state
and a GT$^-$ matrix-element from the final state.
In the $^{54}$Fe and $^{58}$Ni decays,
the $k>2$ configurations are forbidden
in the intermediate GT$^+$ state\cite{ref:NS96},
as far as the initial state is restricted
to the $k\leq 2$ configurations.
Notice that the lowest ({\it i.e.}, $k=0$) configuration
shifts from the parent nucleus to the intermediate one,
because the maximum number of nucleons
which can occupy the $0f_{7/2}$ orbit
changes from nucleus to nucleus in this region.
Unless a configuration has both the GT strengths
from the initial and final states,
this configuration does not contribute
to the $2\nu\beta\beta$ matrix-elements.
Accordingly, only the $k\leq 2$ configurations are relevant
to the $2\nu\beta\beta$ process,
also in the GT$^-$ strength from the final state.
This situation is favorable because the $k\leq 2$ wavefunctions
have been inspected well by spectroscopic studies,
including some single-GT transitions\cite{ref:NS96}.
It should be stated here that these shell-model calculations
in the {\em sd}- and {\em pf}-shell do not provide us
with correct total binding energies.
The Coulomb energy is neglected in the {\em sd}-shell hamiltonian,
while the ground-state energy has been out of scope
in the {\em pf}-shell calculation.
We shift the energies of the intermediate states
so as for the ground-state energy of the intermediate nuclei
to fit to the experimental value,
relative to the initial or final states.
In the $\mbox{$^{36}$Ar}\rightarrow\mbox{$^{36}$S}$
and $\mbox{$^{54}$Fe}\rightarrow\mbox{$^{54}$Cr}$ decays
the modes other than the EC/EC mode cannot occur
because of the $Q$-values,
while the $\beta^+$/EC mode has positive $Q$-value
for $\mbox{$^{58}$Ni}\rightarrow\mbox{$^{58}$Fe}$.

Though the practical application will be devoted to the $\beta^+$ side,
we here discuss how to calculate
the $2\nu\beta\beta$ nuclear matrix-elements more generally.
The nuclear matrix-element of the $2\nu\beta\beta$ process
is written as
\begin{equation} M_\omega(2\nu\beta\beta) = \sum_m
{{\langle f||T({\rm GT}^\pm)||m\rangle
\langle m||T({\rm GT}^\pm)||i\rangle}
\over{E_m-E_i+\omega}} , \label{eq:M_orig} \end{equation}
where $E_i$ and $E_m$ denotes energies
of the initial and intermediate nuclear states, respectively.
The symbol $\omega$ expresses energy of leptons
emitted (or captured) during the step
producing the intermediate state from the initial state.
Its value ranges from zero
to the $Q$-value of the $\beta\beta$ process.
The initial state ($|i\rangle$) is the ground state
of the parent nucleus.
In order for the $\beta\beta$ decays to be detectable,
a single-GT decay from the parent nucleus must be forbidden.
This happens for some decays from even-even nuclei,
owing to the pairing correlation.
Hence the initial states of the $\beta\beta$ decays have $J^P=0^+$.
The intermediate nuclear states ($|m\rangle$),
whose eigenenergy is denoted by $E_m$, should have $J^P=1^+$.
Although the final state ($|f\rangle$) may have $J^P=2^+$
in some $\beta\beta$ decays,
the ground $0^+$ state is expected
to give the dominant branch\cite{ref:DKT85}.
In the practical applications to be discussed,
the final states must be the ground state,
because of the $Q$-values.
The contribution of the Fermi-decay mode due to the isospin-mixing
is known to be negligible\cite{ref:Hax84},
and is omitted in Eq.(\ref{eq:M_orig}).
The bare GT operator is given by
\begin{equation} T^{\rm free}({\rm GT}^\pm)
 = \sum_i g_{A} \sigma_i t_{\pm,i}, \label{eq:GTbare} \end{equation}
where the summation runs over constituent nucleons
and $g_A=G_A/G_V=1.26$.
It has been known that, however,
effects of the core-polarization (CP) and meson-exchange currents (MEC)
should be taken into account in shell-model calculations
on the GT processes\cite{ref:Towner,ref:ASBH}.
For this reason we use effective GT operators,
\begin{equation} T({\rm GT}^\pm) = T^{\rm free}({\rm GT}^\pm)
 + \delta T({\rm GT}^\pm) , \label{eq:GTsp} \end{equation}
where
\begin{equation} \delta T({\rm GT}^\pm)
= \sum_i \left\{ \delta g_{A}(nl) \sigma_i
+ \delta g_{lA}(nl) l_i + \delta g_{pA}(nl, n'l')
 [Y^{(2)}(\hat{\bm{r}}_i) \sigma_i]^{(1)} \right\}
t_{\pm,i} . \label{eq:dGT} \end{equation}
In the shell-model approach,
the summation in Eqs.(\ref{eq:GTbare}) and (\ref{eq:dGT})
is restricted to valence nucleons.
The parameters $\delta g_{A}$, $\delta g_{lA}$ and $\delta g_{pA}$
depend on $n$ and $l$;
the quantum numbers of single-particle orbit
which the $i$-th nucleon occupies.
In the {\em sd}-shell region,
an effective operator for the USD wavefunctions
has been obtained\cite{ref:BW85}, by making a global fitting
of the single-particle parameters of Eq.(\ref{eq:dGT})
to the measured $B({\rm GT})$ values.
In the {\em pf}-shell region,
we adopt the parameter-set evaluated by Towner
from microscopic standpoints\cite{ref:Towner},
which has been shown to yield a sound agreement
with the data\cite{ref:NS96},
combined with the $k\leq 2$ Kuo-Brown wavefunctions.

Application of Eq.(\ref{eq:M_orig})
associated with the above GT operators
to the $2\nu\beta\beta$ matrix-elements
implicates the following two assumptions:
(i) The contribution of high-lying intermediate states
beyond the $0\hbar\omega$ space is negligible,
although a certain amount of the GT strength is carried
out of the $0\hbar\omega$ space due to the CP mechanism.
This is expected because, in addition to large energy-denominator,
the high-lying GT strengths from the initial and final states
are unlikely to be coherent,
and has been supported by a statistical estimate\cite{ref:EEV94}.
(ii) The two sequential GT transitions are separated so well
that the intermediate nuclear states should be
the nuclear energy-eigenstates.
The time interval of the sequential GT transitions
is ruled by the energy denominator via the uncertainty principle,
leading to (1---10MeV)$^{-1}$.
It is much longer than the time-scale of the weak current,
which is the inverse of the gauge-boson mass,
$\sim$(10$^2$GeV)$^{-1}$.
The MEC process takes (10$^2$MeV)$^{-1}$ at most,
owing to the pion mass.
Because the MEC concern the $\delta T({\rm GT})$ term,
the correction due to the time of the meson propagation
will not exceed a few percent
of the total $2\nu\beta\beta$ matrix-elements.
Contribution of other higher-order diagrams
can be estimated to be negligibly small,
in a similar manner.
The assumption (ii) is thus plausible within a few percent accuracy.

The explicit construction of the intermediate energy-eigenstates
($|m\rangle$) in Eq.(\ref{eq:M_orig})
is difficult in most cases,
even if the wavefunctions of the initial and final states
are obtained.
Although the method of solving a coupled linear equation\cite{ref:HO89}
is less elaborate than full diagonalization of a hamiltonian,
still it is not easy to obtain all the possible intermediate states.
This is because there may be
quite a large number of intermediate states.
In the closure approximation adopted
in earlier studies\cite{ref:Hax84},
$E_m$ of Eq.(\ref{eq:M_orig}) is replaced by a constant average value.
We then have, assuming a $0^+$ final state,
\begin{eqnarray} M_\omega(2\nu\beta\beta)
&\sim& {1\over{E_{\rm av}-E_i+\omega}}
\sum_m \langle f||T({\rm GT}^\pm)||m\rangle
\langle m||T({\rm GT}^\pm)||i \rangle \nonumber\\
&=& {1\over{E_{\rm av}-E_i+\omega}}
\langle f||T({\rm GT}^\pm)\cdot T({\rm GT}^\pm)||i \rangle ,
\label{eq:M_clos} \end{eqnarray}
and do not have to handle the intermediate states explicitly.
The average energy $E_{\rm av}$ has been estimated
mainly from the GT$^\pm$ resonance energies.
It has been recognized, both theoretically and experimentally,
that this closure approach hardly gives reliable matrix elements.
This indicates a problem in the estimate of $E_{\rm av}$
and/or in the procedure taking energy average itself.

However, averaging of energy is justified
if we carry it out in a sufficiently small energy range.
Let us consider a set of $1^+$ states
of the intermediate nucleus $\{|\tilde n\rangle\}$
which exhausts the relevant GT$^\pm$ strength from the initial state.
The states $|\tilde n\rangle$'s do not have to be energy eigenstates.
It is possible to insert $\sum_n |\tilde n\rangle\langle\tilde n|$
into the numerator of Eq.(\ref{eq:M_orig}),
\begin{equation} M_\omega(2\nu\beta\beta) = \sum_{m,n}
{{\langle f||T({\rm GT}^\pm)||m\rangle\langle m|\tilde n\rangle
\langle\tilde n||T({\rm GT}^\pm)||i\rangle}
\over{E_m-E_i+\omega}} . \label{eq:M_ins} \end{equation}
It should be commented that the set $\{|\tilde n\rangle\}$
does not have to run out {\it all} the GT strength;
it has only to exhaust the {\it relevant} strength.
For instance,
even if the GT strength has a certain isospin distribution,
only the lowest isospin component needs to be considered
for $\{|\tilde n\rangle\}$,
because higher isospin components do not have
GT-transition strength to $|f\rangle$.
A state $|\tilde n\rangle$ has large overlaps with some eigenstates
and negligibly small ones with others.
These overlaps depend on $E_m$.
It is postulated that the energy distribution of $|\tilde n\rangle$
is peaked so sharply
that $|m\rangle$'s with large $\langle m|\tilde n\rangle$ could have
eigenenergies close to $E_{\tilde n}\equiv
\langle\tilde n|H|\tilde n\rangle$.
We then obtain, via the approximation $E_m\simeq E_{\tilde n}$
under the presence of $\langle m|\tilde n\rangle$,
\begin{equation} M_\omega(2\nu\beta\beta) \simeq \sum_n
{1\over{E_{\tilde n}-E_i+\omega}} \sum_m
\langle f||T({\rm GT}^\pm)||m\rangle\langle m|\tilde n\rangle
\langle\tilde n||T({\rm GT}^\pm)||i \rangle .
\label{eq:M_nout} \end{equation}
The closure with respect to $|m\rangle$ is available at this stage,
giving
\begin{equation} M_\omega(2\nu\beta\beta) \simeq \sum_n
{{\langle f||T({\rm GT}^\pm)||\tilde n\rangle
\langle\tilde n||T({\rm GT}^\pm)||i\rangle}
\over{E_{\tilde n}-E_i+\omega}}. \label{eq:M_app} \end{equation}
The state $|\tilde n\rangle$ plays a role of a doorway state
for a certain number of energy eigenstates $|m\rangle$.
In comparison with Eq.(\ref{eq:M_orig}),
Eq.(\ref{eq:M_app}) appears simply an replacement
of $|m\rangle$ by $|\tilde n\rangle$ and $E_m$ by $E_{\tilde n}$.
Namely, a good approximation will be obtained
by using an appropriate set of doorway states
instead of the intermediate energy-eigenstates.
The approximation of Eq.(\ref{eq:M_app}) can be regarded
as an extension of the closure approximation.
It indeed reduces to a closure approximation
when a single intermediate state (the GT state)
is used for $|\tilde n\rangle$.
The remaining problem is
how to generate $\{|\tilde n\rangle\}$ efficiently.

In Ref.\cite{ref:ZBR90}, a cancellation mechanism has been shown
by using a weakly-coupled two-configuration model.
It is noted that this mechanism is comprehended
in the above equations;
the essential part of the argument in Ref.\cite{ref:ZBR90} reappears
if we regard $|\tilde n\rangle$ as a state
having a considerable GT$^\pm$ strength from $|i\rangle$
but no GT$^\mp$ strength from $|f\rangle$.
The vanishing GT$^\mp$ strength prohibits $|\tilde n\rangle$
from contributing to $M_\omega(2\nu\beta\beta)$
(see Eq.(\ref{eq:M_app})).
There may be one or more states with opposite character, 
having a GT$^\mp$ strength from $|f\rangle$
but no GT$^\pm$ strength from $|i\rangle$.
Suppose that some of these states
have energies close to $E_{\tilde n}$,
and couple weakly to $|\tilde n\rangle$.
This coupling gives rise to a certain fragmentation
of $|\tilde n\rangle$
over several eigenstates,
which is here denoted by $|m_1\rangle$, $|m_2\rangle$,
$\cdots$, $|m_r\rangle$.
Their eigenenergies are close to $E_{\tilde n}$.
Each of $|m_1\rangle$, $|m_2\rangle$, $\cdots$, $|m_r\rangle$
may have a sizable contribution to $M_\omega(2\nu\beta\beta)$,
because they have GT strengths both from $|i\rangle$ and $|f\rangle$.
These contributions, however, cancel one another,
and the result obtained solely by $|\tilde n\rangle$,
which gives the vanishing contribution to $M_\omega(2\nu\beta\beta)$,
is recovered.
The whole contribution of $|m_1\rangle$, $|m_2\rangle$, $\cdots$,
$|m_r\rangle$ to $M_\omega(2\nu\beta\beta)$
is represented by $|\tilde n\rangle$,
as is shown from Eq.(\ref{eq:M_ins}) to (\ref{eq:M_app})
in more general cases.

The set $\{|\tilde n\rangle\}$ has been assumed
to exhaust the transition strength from the state $|i\rangle$,
in the above discussion leading to Eq.(\ref{eq:M_app}).
It should be noted that
an analogous equation is derived by adopting a set of doorway states
so as to exhaust the relevant GT strengths from $|f\rangle$.
These two ways give the same value if they are convergent,
although the speed of convergence may be different.

The present derivation of Eq.(\ref{eq:M_app}) discloses
what conditions are required for $\{|\tilde n\rangle\}$.
There are two conditions:
(a) The set $\{|\tilde n\rangle\}$ must exhaust
the relevant GT strength
either from the initial or final state,
and (b) each doorway state $|\tilde n\rangle$
has an energy distribution over $|m\rangle$
with sharp peak around $E_{\tilde n}$,
so as for $E_m\simeq E_{\tilde n}$ in the energy denominator
to be justified.
A promising way to obtain such approximate intermediate states
has been proposed in Ref.\cite{ref:CPZ90},
by applying Whitehead's moment method\cite{ref:Wh80}.
The algorithm is as follows:
\begin{enumerate}
\item Produce the state exhausting the relevant GT$^\pm$ strength
from the initial (final) state,
$|{\rm GT}^\pm_{i(f)}\rangle \propto P_{\rm rel}\cdot
 T({\rm GT}^\pm)|i(f)\rangle$,
where $P_{\rm rel}$ stands for a projection operator
which picks up the relevant configurations.
\label{it:GT}
\item Generate a set of bases
by operating the shell-model hamiltonian $H$
on $|{\rm GT}^\pm_{i(f)}\rangle$;
$H|{\rm GT}^\pm_{i(f)}\rangle$, $H^2|{\rm GT}^\pm_{i(f)}\rangle$,
$\cdots$, $H^N|{\rm GT}^\pm_{i(f)}\rangle$.
\label{it:base}
\item Diagonalize $H$ within the space
$\Gamma^{(N)}\equiv\left\{|{\rm GT}^\pm_{i(f)}\rangle,
H|{\rm GT}^\pm_{i(f)}\rangle, \cdots,
H^N|{\rm GT}^\pm_{i(f)}\rangle\right\}$.
Each energy eigenvalue and eigenstate in this subspace
are taken as $E_{\tilde n}$ and $|\tilde n\rangle$.
\label{it:diag}
\item Increase $N$, until $M_\omega(2\nu\beta\beta)$ converges.
\label{it:iter}
\end{enumerate}
Although the steps (\ref{it:base}) and (\ref{it:diag})
are common with the Lanczos diagonalization procedure,
the convergence of $M_\omega(2\nu\beta\beta)$ does not necessarily
imply the convergence about the intermediate states,
as will be shown later.
The condition (a) is obviously fulfilled,
since $\Gamma^{(N)}$ contains $|{\rm GT}^\pm_{i(f)}\rangle$,
into which all the relevant GT strength is concentrated.
The simplest case of $N=1$ is equivalent
to the closure approximation of Ref.\cite{ref:Hax84}.
Speed of the convergence for $N$ depends on how well (b) is satisfied.
This point will be examined below for the practical cases.

Since all the $\beta\beta$ decays to be investigated
belong to the $\beta^+$ side,
the present practical calculation has to do
only with $T({\rm GT}^+)$, not $T({\rm GT}^-)$,
in Eqs.(\ref{eq:M_orig}---\ref{eq:M_app}).
For the GT state generated from the initial or final state
at the step (\ref{it:GT}),
$|{\rm GT}^+_i\rangle$ or $|{\rm GT}^-_f\rangle$ is considered.
When $|{\rm GT}^+_i\rangle$ is chosen at (\ref{it:GT}),
we do not use the $P_{\rm rel}$ operator
({\it i.e.}, $P_{\rm rel}=1$).
For $|{\rm GT}^-_f\rangle$,
$P_{\rm rel}$ represents the isospin projection
for the $^{36}$Ar decay,
while for the $^{54}$Fe and $^{58}$Ni decays
the projection onto the $k\leq 2$ space is included as well.

Because of the phase space of the emitted leptons,
$\omega\sim Q/2$ yields a main contribution
to the $\beta\beta$ probabilities,
where $Q$ denotes the $Q$-value of the $\beta\beta$ process.
We examine convergence of the above method,
by taking $\omega=Q/2$ as a typical value of $\omega$.
Fig.\ref{fig:conv} shows the convergence
for $N$ (the number of the iteration).
The result on the EC/EC mode is presented for the $^{58}$Ni decay.
The $\beta^+$/EC mode has no essential difference.
The number of the $J^P=1^+, T=1$ ({\it i.e.}, relevant) states
with the $sd$-shell configuration is 54 in $^{36}$Cl,
the intermediate product of the $^{36}$Ar decay.
Hence a full diagonalization of the shell-model hamiltonian 
is possible.
On the other hand, for $^{54}$Mn and $^{58}$Co,
there are more than $10^4$ states
having $J^P=1^+$ and the lowest isospin,
even in the present truncated configuration space.
Nevertheless, the convergence is so rapid
that $N=10$ could be almost sufficient for all these cases.
This situation is quite similar to the $^{48}$Ca decay
reported in Ref.\cite{ref:CPZ90}.
The cases starting from the GT$^+$ state
and the GT$^-$ state are compared for the $^{36}$Ar decay.
We view that the GT$^+$ case yields somewhat better convergence,
though the difference is not quite big.
The same holds for the $^{54}$Fe and $^{58}$Ni decays.
The reason for this will be discussed later.
\begin{figure}
\begin{center}
\vspace{30mm}
\caption{\label{fig:conv}}
\end{center}
\end{figure}

In Fig.\ref{fig:Fe54-GT+},
the GT$^+$ strengths from $^{54}$Fe are depicted
for $N=10$, $30$ and $150$.
Obviously the strength function does not converge
with $N=10$ and $30$.
The convergence of the GT$^-$ strength function
is somewhat worse than the GT$^+$ one.
Nevertheless, $M_\omega(2\nu\beta\beta)$ is convergent,
as shown in Fig.\ref{fig:conv}.
This is because the condition (b) for the set $\{|{\tilde n}\rangle\}$
is satisfied well, even with such a small number for $N$.
We can track how each strength for $N=10$ diffuses as $N$ increases.
For instance, about 90\% of the GT$^+$ strength
at $E_x=2.7$MeV appearing in the $N=10$ case
remains in $E_x=2.0$---$3.1$MeV in the $N=150$ result.
For the strengths with $N=30$,
about $90\%$ of the $E_x=2.9$MeV strength is shared
only by the states at $E_x=2.7$ and $3.1$MeV of the $N=150$ result.
In this manner,
the $N=10$ or $30$ strengths, particularly the low-lying ones,
are found to stay in a small energy range
around the original position.
It is confirmed that the $E_x\lsim 5$MeV strengths are convergent
with $N=150$,
whereas higher-lying ones are not.
\begin{figure}
\begin{center}
\vspace{30mm}
\caption{\label{fig:Fe54-GT+}}
\end{center}
\end{figure}

\section{Dominance of low-lying states in intermediate nucleus}
\label{low-lying}

We next investigate,
taking notice of the excitation energies,
which intermediate states
give significant contributions
to the $2\nu\beta\beta$ matrix-elements.

Fig.\ref{fig:Edist} shows contribution
of the intermediate $1^+$ states
to $M_{\omega=Q/2}(2\nu\beta\beta)$,
as a function of the excitation energy.
Summed value of the $2\nu\beta\beta$ matrix-elements
up to the excitation energy $E_x$,
\begin{displaymath} \sum_{n~{\rm for}~E_{\tilde n}-E_0<E_x}
{{\langle f||T({\rm GT}^\pm)||\tilde n\rangle
\langle\tilde n||T({\rm GT}^\pm)||i\rangle}
\over{E_{\tilde n}-E_i+\omega}}, \end{displaymath}
is displayed,
in order to make the convergence for the energy transparent.
The ground-state energy of the intermediate nucleus
is denoted by $E_0$ here.
Thus the contribution of each intermediate state
is presented by the stepwise increase or decrease.
The result of the full diagonalization is shown
for the $^{36}$Ar decay,
while the $N=150$ results for the $^{54}$Fe and $^{58}$Ni decays.
The computation of $M_\omega(2\nu\beta\beta)$
begins with $|{\rm GT}^+_i\rangle$ at the step (\ref{it:GT})
(see Section~\ref{sec:comp}).
It is found that low-lying intermediate states
dominate $M_\omega(2\nu\beta\beta)$.
The $E_x\lsim 5$MeV intermediate states govern
$M_\omega(2\nu\beta\beta)$ for the $^{54}$Fe and $^{58}$Ni cases,
while a few of the $5<E_x\lsim 10$MeV states
give a sizable contribution for $^{36}$Ar.
This difference in energy might be caused
by the difference in mass region.
Furthermore, the number of the intermediate states
giving considerable contribution to $M_\omega(2\nu\beta\beta)$
is not very large.
The dominance of a relatively small number of low-lying states
in $M_\omega(2\nu\beta\beta)$
seems consistent with the suggestion
in Ref.\cite{ref:ZBR90} for the $^{48}$Ca decay,
although the lowest $1^+$ state is not enough
to approximate $M_\omega(2\nu\beta\beta)$ in the present cases.
\begin{figure}
\begin{center}
\vspace{30mm}
\caption{\label{fig:Edist}}
\end{center}
\end{figure}

We shall look into
the $\mbox{$^{36}$Ar}\rightarrow\mbox{$^{36}$S}$ case
in more detail.
The shell-model hamiltonian is fully diagonalized,
without any further truncation in the $sd$-shell.
The energy distribution of the contribution
to the EC/EC nuclear matrix-element
has been shown in Fig.\ref{fig:Edist},
being repeated in the middle sector of Fig.\ref{fig:Ar}.
The upper (lower) sector of  Fig.\ref{fig:Ar} displays
the energy distribution of the GT$^+$ (GT$^-$ with $T=1$) strength
from the initial (final) state.
The GT$^+$ strength distribution correlates rather well
to the steps of increase or decrease
in the graph of $M_\omega(2\nu\beta\beta)$,
though the size of the steps relative to the GT$^+$ strengths
comes smaller as the energy increases.
The distribution of the $T=1$ GT$^-$ strength
is quite different from that of GT$^+$,
having greater weights in high-energy region.
Indeed, the overlap between $|{\rm GT}^+_i\rangle$
and $|{\rm GT}^-_f\rangle$
is very small ($\sim 2\%$).
Higher-lying GT$^-$ strengths hardly contribute
to $M_\omega(2\nu\beta\beta)$,
on account of the small overlap to the GT$^+$ strength
as well as of the large energy-denominator.
Only a small portion (low-lying part) of the GT$^-$ strength
contributes to the $2\nu\beta\beta$ matrix-element.
This fact accounts for the slower convergence
from $|{\rm GT}^-_f\rangle$ than from $|{\rm GT}^+_i\rangle$.
This mechanism applies also to the $^{54}$Fe and $^{58}$Ni decays.
\begin{figure}
\begin{center}
\vspace{30mm}
\caption{\label{fig:Ar}}
\end{center}
\end{figure}

Contribution of the low-lying intermediate states
is enhanced to a certain extent,
due to the energy denominator.
The effect of the energy denominator is, however,
insufficient to account for the remarkable dominance
of the low-lying states clarified presently.
The particular significance of the low-lying GT strengths
is qualitatively understood from a general argument
on the shell structure of nuclei.
To the first approximation,
the initial and final states have configurations
depicted by Fig.\ref{fig:config}--a) and b),
because they are lowest-lying.
The boxes in the graphs indicate the orbits
occupied commonly between the initial and final states.
The GT$^+$ transition from the configuration
of Fig.\ref{fig:config}--a)
generates the configurations of Fig.\ref{fig:config}--c),
while Fig.\ref{fig:config}--b) leads to Fig.\ref{fig:config}--d).
The $2\nu\beta\beta$ decay proceeds through the common configuration
between Fig.\ref{fig:config}--c) and d).
It is immediately found out that only the first graphs,
which actually imply the low-energy part
of the GT$^+$ and GT$^-$ strengths,
are common between c) and d).
Therefore, contribution of higher-lying GT strengths
is hindered by the shell structure,
as well as by the energy denominator.
This shell effect is ignored in the closure approximation,
making the energy denominator too large
and giving rise to an underestimate of $M_\omega(2\nu\beta\beta)$,
as viewed in the $N=1$ limits of Fig.\ref{fig:conv}.
It is also noticed that, compared with the GT$^+$ transition,
the GT$^-$ one has more patterns of high-lying configurations,
originating in the neutron excess.
The central energy of the GT$^-$ strength
is therefore higher than that of GT$^+$,
as has been seen in Fig.\ref{fig:Ar},
even if we restrict ourselves to the lowest isospin component.
\begin{figure}
\begin{center}
\vspace{30mm}
\caption{\label{fig:config}}
\end{center}
\end{figure}

Let us consider the shell-structure effect
on $M_\omega(2\nu\beta\beta)$ more concretely,
for the example of the $^{36}$Ar decay.
The lowest configurations of $^{36}$Ar and $^{36}$S are
\begin{eqnarray} (p: 0d_{3/2}^{-2},~n: 0d_{3/2}^{-2})
&&\mbox{for $^{36}$Ar}, \label{eq:Ar_config} \\
(p: 0d_{3/2}^{-4})&&\mbox{for $^{36}$S},
\label{eq:S_config} \end{eqnarray}
on top of the $^{40}$Ca core.
Thereby the GT$^+$ excitation from $^{36}$Ar
and the GT$^-$ excitation from $^{36}$S primarily produce
the following configurations of $^{36}$Cl,
\begin{eqnarray} (p: 0d_{3/2}^{-3},~n: 0d_{3/2}^{-1}),
~~(p: 0d_{3/2}^{-2} 0d_{5/2}^{-1},~n: 0d_{3/2}^{-1})
&&\mbox{for GT$^+$ from $^{36}$Ar}, \nonumber\\
\\
(p: 0d_{3/2}^{-3},~n: 0d_{3/2}^{-1}),
~~(p: 0d_{3/2}^{-3},~n: 0d_{5/2}^{-1})
&&\mbox{for GT$^-$ from $^{36}$S} . \nonumber\\
\end{eqnarray}
The common configuration $(p: 0d_{3/2}^{-3},~n: 0d_{3/2}^{-1})$,
which is the lowest configuration of $^{36}$Cl,
gives the main contribution to the $2\nu\beta\beta$ matrix-element.
The $(p: 0d_{3/2}^{-3},~n: 0d_{3/2}^{-1})$ constitutes
only a single $1^+$ state with $T=1$.
This component is, however, fragmented in real situations,
because other $0\hbar\omega$ configurations
appreciably mixes due to the residual interaction.
Together with the leakage out of the configurations
of Eqs.(\ref{eq:Ar_config},\ref{eq:S_config})
for the initial and final states,
this fragmentation of the $1^+$ component
distributes the $2\nu\beta\beta$ matrix-elements
over a certain range of energy.
It is noted that
the $(p: 0d_{3/2}^{-3},~n: 0d_{5/2}^{-1})$ configuration
yields higher energy than
the $(p: 0d_{3/2}^{-2} 0d_{5/2}^{-1},~n: 0d_{3/2}^{-1})$ one,
because the neutrons occupying $0d_{5/2}$
are harder to be excited than the protons in the same orbit.
In terms of the BCS theory,
the quasiparticle energy of $0d_{5/2}$ is higher for neutrons
than for protons.
Hence the GT$^-$ energy should be substantially higher
than the GT$^+$ one.

In the $^{54}$Fe and $^{58}$Ni cases,
the $2\nu\beta\beta$ processes would be forbidden,
if merely $1p_{3/2}$ were active for neutrons.
The principal contribution comes from the configurations
with the excitation from $1p_{3/2}$ to $0f_{5/2}$.
It still holds that
$2\nu\beta\beta$ proceeds mainly through lower configurations
of the intermediate nucleus.
Thus the low-lying intermediate states
dominate the $2\nu\beta\beta$ matrix-elements,
as shown in Fig.\ref{fig:Edist}.

We here mention a speciality
of the $\mbox{$^{48}$Ca}\rightarrow\mbox{$^{48}$Ti}$ decay.
To the first approximation,
$^{48}$Ca can be regarded as a doubly magic core.
The lowest configuration of $^{48}$Ti is
$(p: 0f_{7/2}^{~2},~n: 0f_{7/2}^{-2})$.
The $2\nu\beta\beta$ decay is mediated
by the $(p: 0f_{7/2}^{~1},~n: 0f_{7/2}^{-1})$ configuration,
which is the lowest configuration of $^{48}$Sc.
This configuration allows a single $1^+$ state,
like the $^{36}$Ar case.
On the other hand, the $0f_{7/2}$ orbit is isolated
and the residual interaction can hardly mix this configuration
with other ones.
Therefore it remains almost pure,
constituting the lowest $1^+$ state.
This consideration accounts for the reason
that the lowest $1^+$ state already gives a good approximation
to the total $2\nu\beta\beta$ matrix-element
in the $^{48}$Ca decay, as pointed out in Ref.\cite{ref:ZBR90},
unlike the other cases investigated in this article.
The dominance of the lowest $1^+$ state has been suggested
from experiments for the $^{100}$Mo decay\cite{ref:Mo100},
and an extensive study has been carried out
also for several other decays\cite{ref:ET96}.
This can also be explained by a similar argument
for the nuclei with $Z$ or $N=40-50$,
in which the isolation of the $0g_{9/2}$ orbit
plays an essential role.
The $^{128}$Te and $^{130}$Te cases\cite{ref:Mo100,ref:ET96}
are a different matter.
Another effect such as quadrupole deformation might be necessary
to understand this case.

The role of the shell structure has not been recognized
sufficiently so far.
Despite the arguments for the $^{48}$Ca decay\cite{ref:ZBR90},
there has been a lack of such discussions
adapting to more general cases.
In the $0\nu\beta\beta$ decays,
the difference of energy denominator
among low- and high-lying intermediate states
is much less important\cite{ref:DKT85} than in $2\nu\beta\beta$.
However, the shell-structure mechanism remains
and lower-lying GT strengths will still be significant,
because of a certain similarity
of the $0\nu\beta\beta$ nuclear matrix-elements
to the $2\nu\beta\beta$ ones.

The present calculation has been carried out
with the algorithm developed in Ref.\cite{ref:CPZ90}.
It has been confirmed extensively in this article
how powerful this method is,
by taking examples other than the $^{48}$Ca decay
as well as by a detailed discussion. 
There are a few other minor differences
of the present study from that of Ref.\cite{ref:CPZ90}.
While the GT operator proportional to $\sigma t_\pm$
has been used in Ref.\cite{ref:CPZ90},
more realistic GT operators are adopted here.
This does not influence the speed of convergence, as expected.
Although only the computation starting from $|{\rm GT}^-_i\rangle$
is considered in Ref.\cite{ref:CPZ90},
we have tested both the $|{\rm GT}^+_i\rangle$
and $|{\rm GT}^-_f\rangle$ cases.
A slight faster convergence of the $|{\rm GT}^+_i\rangle$ case
has become evident.
It is naturally expected in the $\beta^-\beta^-$ decays
that the calculation starting from $|{\rm GT}^+_f\rangle$
converges somewhat faster than from $|{\rm GT}^-_i\rangle$.

The quasiparticle random-phase approximation (QRPA)
has often been applied to calculating
the $\beta\beta$ nuclear matrix-elements.
It is known, however,
that results of the $2\nu\beta\beta$ calculation with QRPA
are seriously sensitive to $g_{pp}$
around $g_{pp}=1$\cite{ref:gpp,ref:MBK89},
where $g_{pp}$ denotes a factor
introduced for the particle-particle interaction.
This problem regarding $g_{pp}$ is crucial
to predictability of the QRPA calculation.
It has been shown in Ref.\cite{ref:MBK89}
that the low-lying GT spectra, particularly the GT$^+$ ones,
are sensitive to $g_{pp}$.
Based on the above discussion of the shell-structure effect
and the consequence of the present shell-model calculations,
the $g_{pp}$ problem in the $2\nu\beta\beta$ calculation
seems to originate primarily in the sensitivity
of the low-lying GT$^+$ strengths to $g_{pp}$.

It is emphasized that the low-lying GT strengths
with $E_x\lsim\mbox{(5---10)MeV}$,
rather than higher-lying component such as the GT$^-$-resonance,
are essential to the $2\nu\beta\beta$ nuclear matrix-elements.
There are several significant implications.
As is well-known, energies and wavefunctions of lower-lying states
converge rapidly in the Lanczos diagonalization algorithm.
Convergence of the present algorithm is accelerated further
by the dominance of the low-lying GT strengths,
because its essential part is the same as the Lanczos method.
The low-lying GT strengths will provide us with a crucial test
on reliability of the calculations
of the $2\nu\beta\beta$ nuclear matrix-elements.
It is true also for the QRPA calculations.
Both of the parent and daughter nuclei of the $\beta\beta$ decays
are stable against the single-$\beta$ decay.
It is hopeful to measure
by high-resolution charge-exchange experiments
the low-lying GT strengths from the initial and final states. 
Theoretical approaches should be examined critically
for the low-lying GT strengths by these measurements.
As has been discussed in connection with Eq.(\ref{eq:M_app}),
concise reproduction of individual GT strength
is not necessarily required.
It is requisite, however, to describe properly
the average character of the low-lying strengths
in a sufficiently small energy range.
This point is more important
than to reproduce the total GT strengths.
Moreover, if we had a reliable assessment
on the signs of the GT matrix-elements,
the $2\nu\beta\beta$ matrix-element could be obtained experimentally
by using the measured GT strengths.
In the present calculations and those for the $^{48}$Ca decay,
energy regions adding positive and negative values
to $M_\omega(2\nu\beta\beta)$ appear nearly separated.
It may not be difficult to learn the signs
from realistic calculations.

\section{Half-lives of $2\nu\beta\beta$ decays}
\label{sec:half-life}

The main purpose of this article is to clarify
what is required to calculate
the $2\nu\beta\beta$ nuclear matrix-elements.
As far as the calculated GT strengths
from the initial and final states
have not been examined sufficiently,
we should not claim high precision
for the $2\nu\beta\beta$ matrix-elements
calculated in the present study.
We here show, however, the calculated half-lives,
because there could be some interest in the prediction
of the present realistic shell-model calculations.
In order to evaluate half-lives of the $2\nu\beta\beta$ decays,
integration of the lepton degrees-of-freedom is carried out,
by following Ref.\cite{ref:DK92}.
As a result, half-lives of $T^{2\nu}_{1/2}({\rm EC}/{\rm EC};
\mbox{$^{36}$Ar}\rightarrow\mbox{$^{36}$S})
=1.7\times 10^{29}\mbox{yr}$,
$T^{2\nu}_{1/2}({\rm EC}/{\rm EC};
\mbox{$^{54}$Fe}\rightarrow\mbox{$^{54}$Cr})
=1.5\times 10^{27}\mbox{yr}$
are obtained from the present nuclear matrix-elements.
The half-lives of the $^{58}$Ni decay is shown
in Table~\ref{tab:Ni58},
in comparison with the QRPA results\cite{ref:HMOK}
and the data\cite{ref:Ni58}.
The present calculation gives half-lives somewhat longer,
but comparable to those by the QRPA calculation.
Both calculations predict much longer lifetimes
than the lower-limits of experiments.

\section{Summary}
\label{sec:summary}

It has been discussed what conditions are demanded
to compute the $2\nu\beta\beta$ nuclear matrix-elements:
We should prepare for a set of (approximate) intermediate states
with (a) exhausting the relevant GT strength
and (b) energy distribution having a sharp peak,
over the energy eigenstates.
An algorithm proposed in Ref.\cite{ref:CPZ90}
based on Whitehead's moment method seems quite promising
from this viewpoint,
particularly when it is associated
with large-scale realistic shell-model wavefunctions.
This algorithm has been inspected,
applied to the $\mbox{$^{36}$Ar}\rightarrow\mbox{$^{36}$S}$,
$\mbox{$^{54}$Fe}\rightarrow\mbox{$^{54}$Cr}$ and
$\mbox{$^{58}$Ni}\rightarrow\mbox{$^{58}$Fe}$ decays.
Quite rapid convergence has been shown.
It has also been pointed out that
the low-lying part of the intermediate GT strength
dominates the $2\nu\beta\beta$ rates,
owing to the shell structure of nuclei
as well as to the energy denominator.
However, it will not be always true
that only the lowest intermediate $1^+$ state
is enough in evaluating the $2\nu\beta\beta$ nuclear matrix-elements.
It somewhat depends on the extent of the configuration mixing.
Careful tests of the low-lying GT strengths,
which will be doable with assistance
of high-resolution charge-exchange experiments,
are significant to predict the $2\nu\beta\beta$ matrix-elements.
Half-lives of those decays
obtained within the present shell-model framework
have been reported.

$\\$
The authors are grateful to Dr. S. Cohen
for careful reading of the manuscript.
The numerical calculation is financially supported
by Research Center for Nuclear Physics, Osaka University
(Project No.95-A-01).

\clearpage
\pagestyle{empty}
\section*{Figure Captions}
\begin{description}
\item[Fig.\ref{fig:conv}:]
Convergence of $M_{\omega=Q/2}(2\nu\beta\beta)$ for $N$.
The upper sector shows the convergence
starting from $|{\rm GT}^+_i\rangle$
for the three decays.
In the lower sector,
the convergence starting from $|{\rm GT}^+_i\rangle$
and that from $|{\rm GT}^-_f\rangle$
are compared for the $^{36}$Ar decay.
\item[Fig.\ref{fig:Fe54-GT+}:]
GT$^+$ strength function for $^{54}$Fe
calculated by Whitehead's moment method:
the strength functions obtained with $N=10$ (upper),
$30$ (middle) and $150$ (lower) are compared.
\item[Fig.\ref{fig:Edist}:]
$M_{\omega=Q/2}(2\nu\beta\beta)$
with contribution of the intermediate states
up to the excitation energy specified by the horizontal axis.
\item[Fig.\ref{fig:Ar}:]
GT$^+$ strength function of $^{36}$Ar (upper),
GT$^-$ strength function of $^{36}$S (lower)
and $M_{\omega=Q/2}(2\nu\beta\beta)$ up to $E_x$
for the $^{36}$Ar decay (middle).
\item[Fig.\ref{fig:config}:]
Proton and neutron schematic configurations
regarding the $\beta\beta$ decay.\protect\\
a,b): Ground-state configurations
of the initial and final nuclei.\protect\\
c) Possible configurations generated by the GT$^+$ transition
from the state of a).\protect\\
d) Possible configurations generated by the GT$^-$ transition
from the state of b).
\end{description}

\clearpage
\section*{Tables}
\begin{table}[htb]
\begin{center}
\begin{small}
\caption{\label{tab:Ni58}
Half-lives (yr) of
the $\mbox{$^{58}$Ni}\rightarrow\mbox{$^{58}$Fe}$ decay.
The values obtained in the present calculation are compared
with those of the QRPA calculation of Ref.\protect\cite{ref:HMOK}
and of the experimental lower-limits\protect\cite{ref:Ni58}.}
\begin{tabular}{cr@{$\times$}lr@{$\times$}lr@{$\times$}l}
   \hline
     & \multicolumn{2}{c}{Present work} & \multicolumn{2}{c}{QRPA}
     & \multicolumn{2}{c}{Exp.} \\
   \hline
     EC/EC &~~~$6.1$&$10^{24}$~~~&~~~$3.9$&$10^{24}$~~~
     &~~~$>2.1$&$10^{19}$~~~\\
     $\beta^+$/EC & $8.6$&$10^{25}$ & $5.5$&$10^{25}$
     & $>6.2$&$10^{19}$ \\
   \hline
\end{tabular}
\end{small}
\end{center}
\end{table}

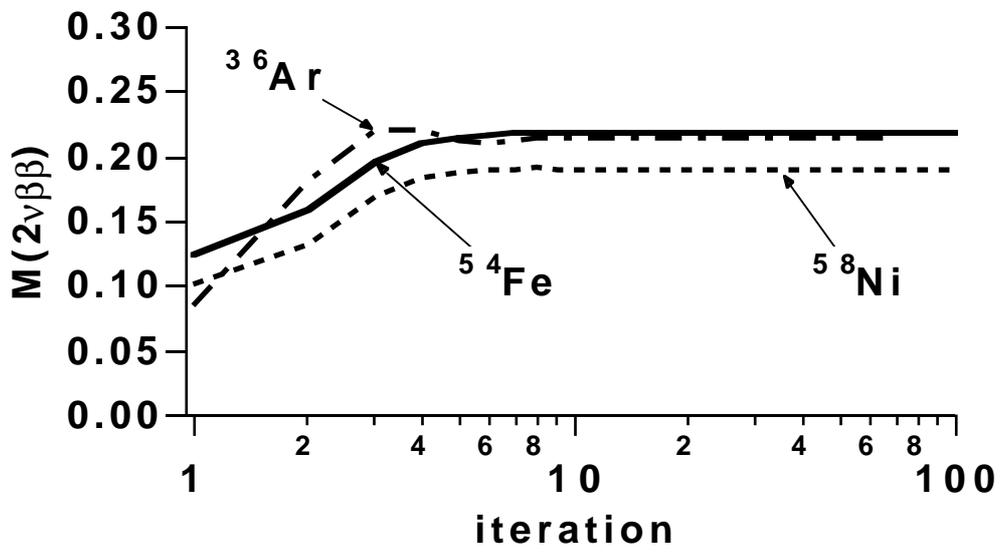
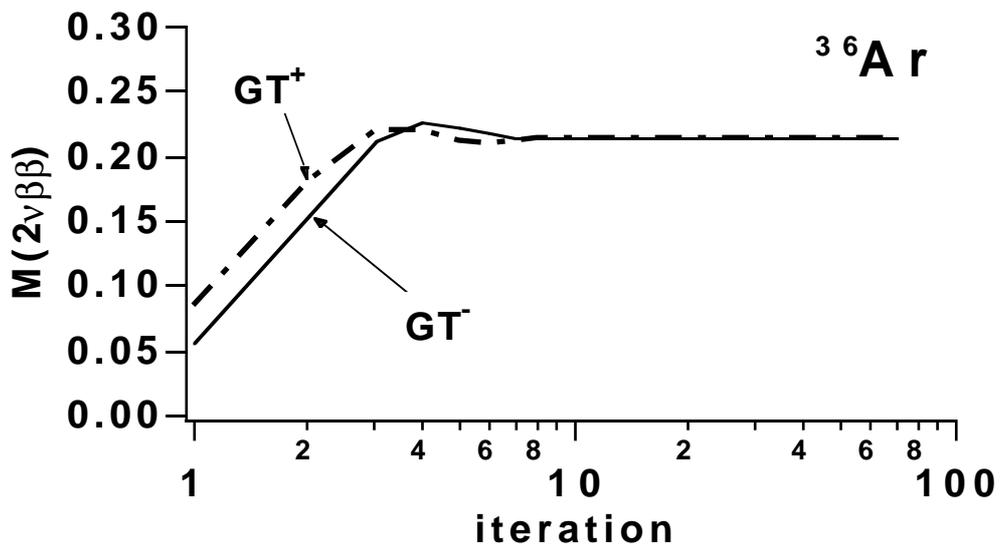
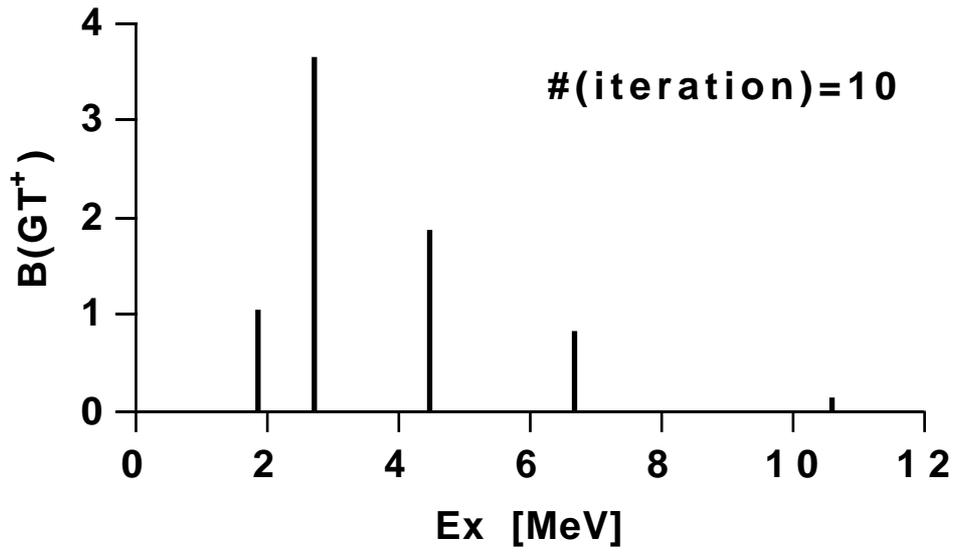
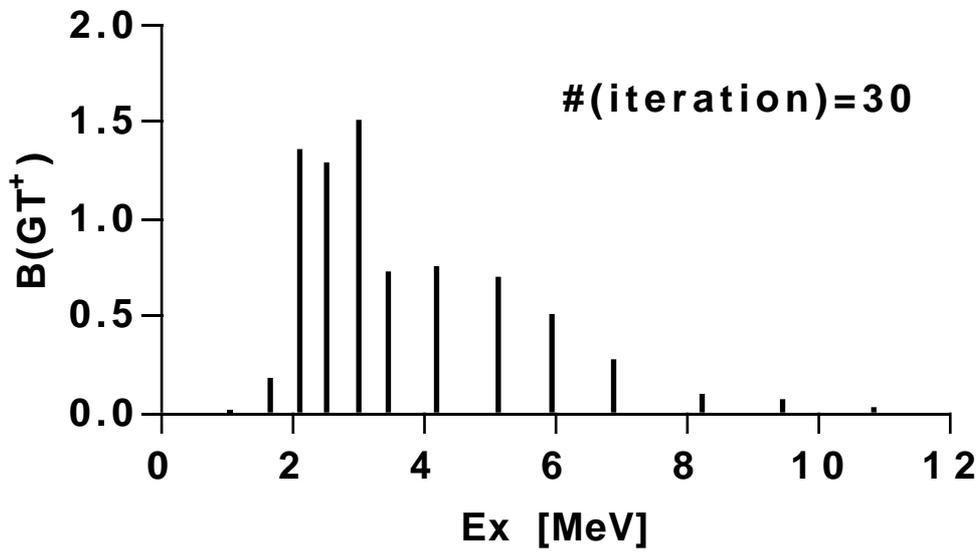
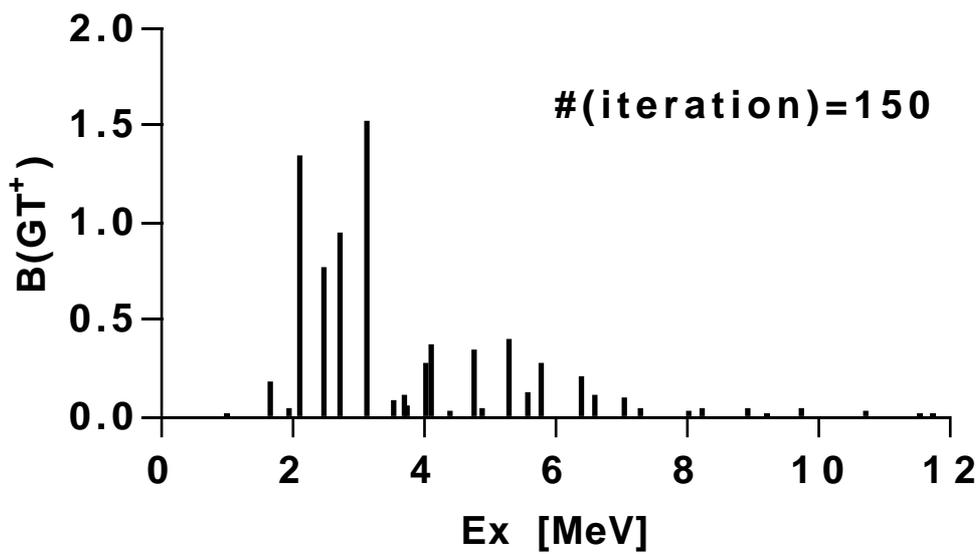
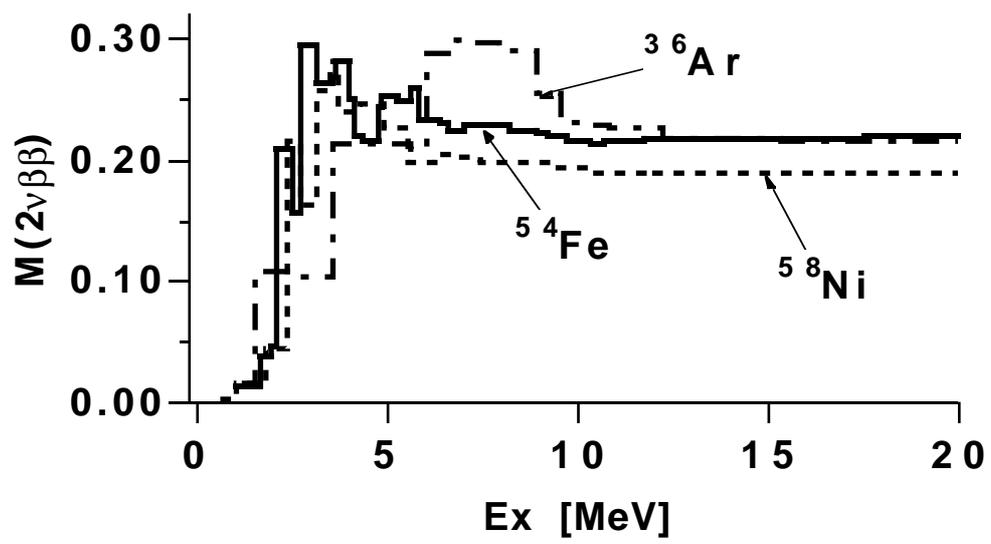
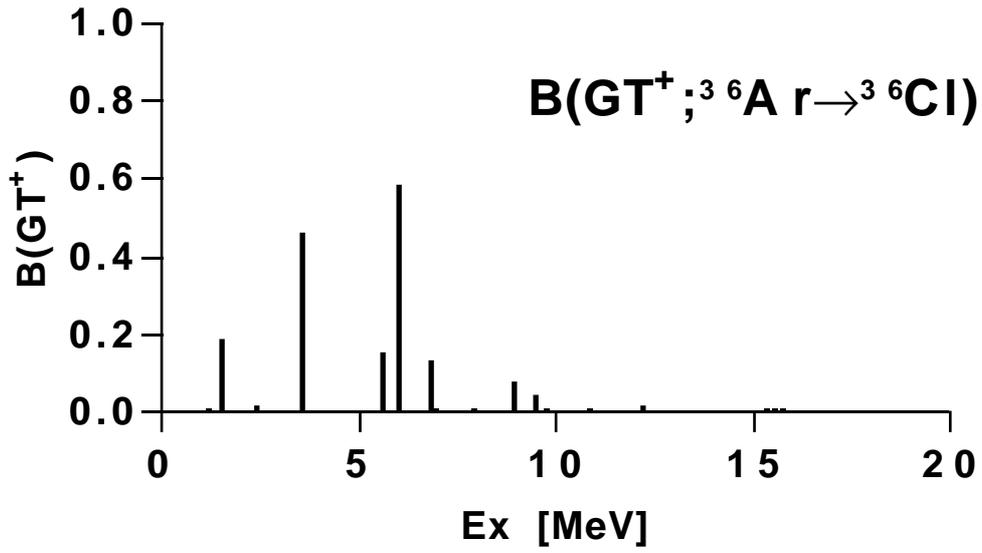
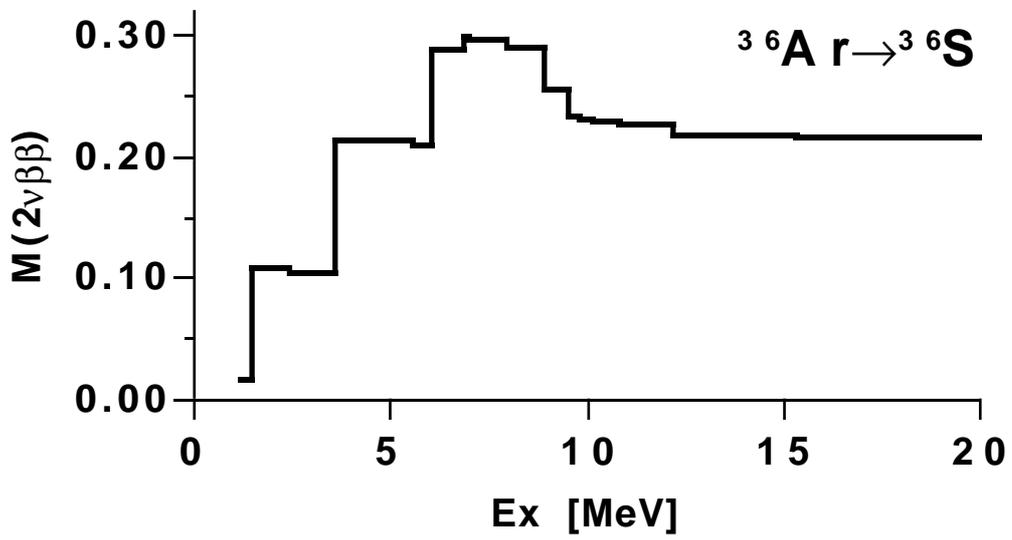
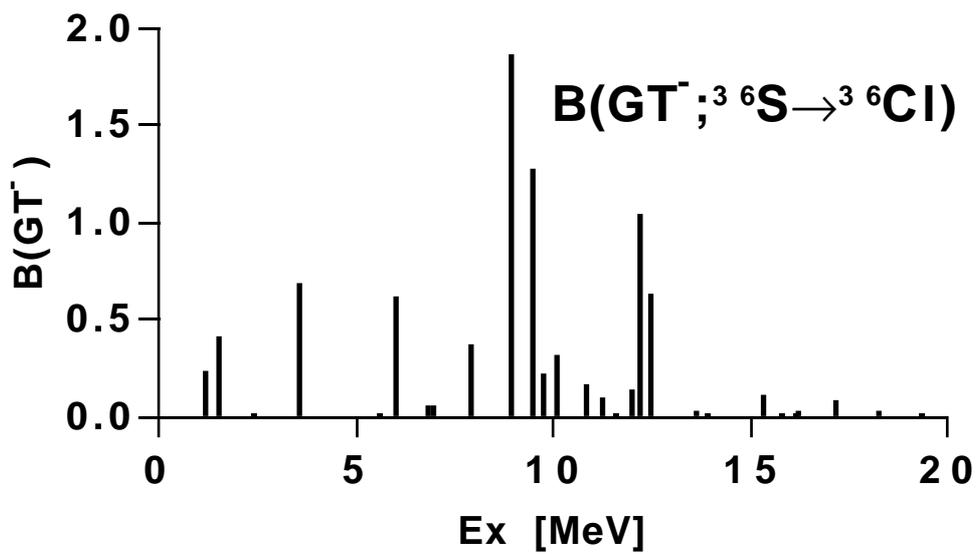
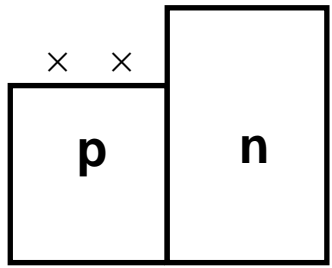
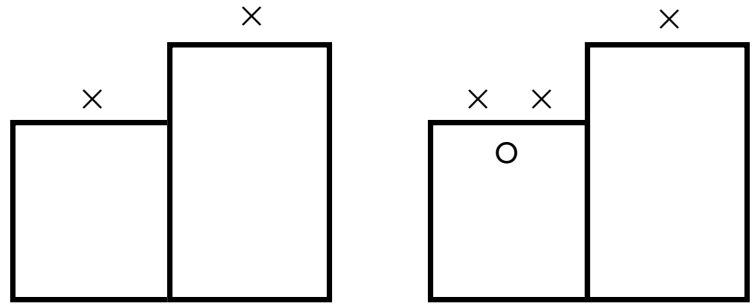
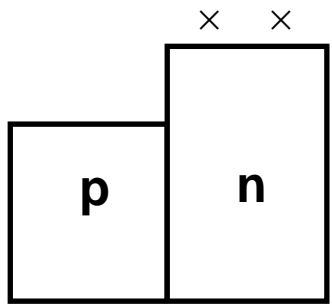
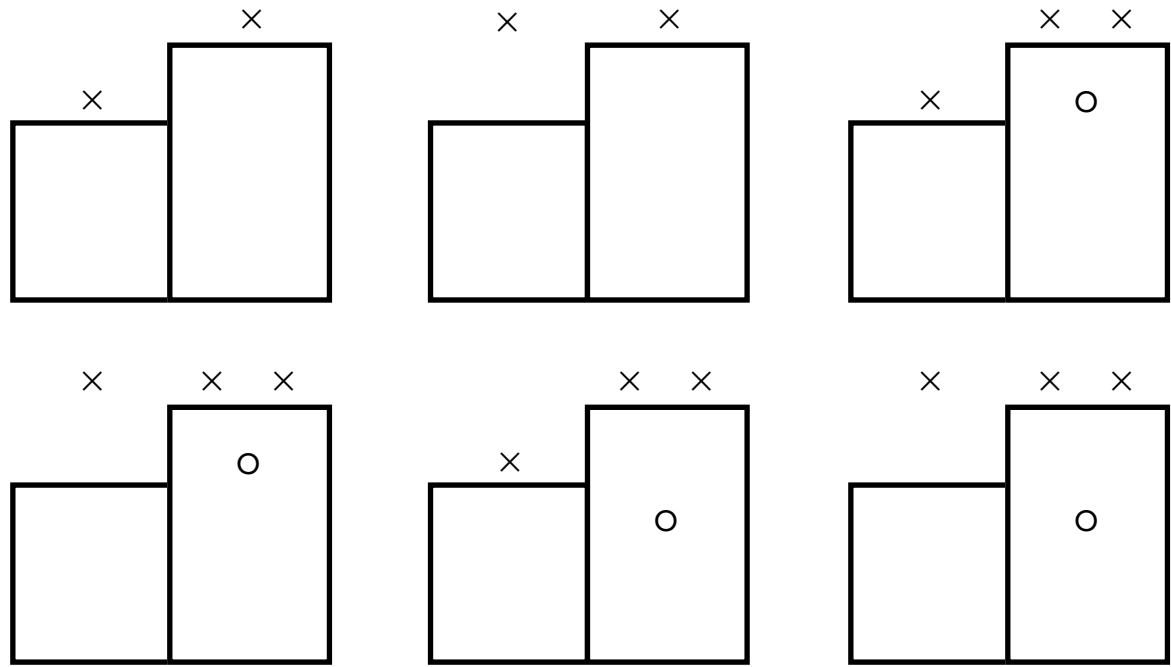
\end{document}